\documentclass[12pt]{article}
\usepackage{graphicx}
\usepackage{amsmath}

\usepackage{amscd,amssymb}
\usepackage{hyperref}
\usepackage[cmtip,arrow,
matrix]{xy}

\usepackage{curves}

\def\cR{{\mathcal R}}

\def\cN{{\mathcal N}}
\def\cA{{\mathcal A}}

\def\cC{{\mathcal C}}
\def\cK{{\mathcal K}}

\def\cH{{\mathcal H}}

\def\cQ{{\mathcal Q}}

\begin{document}

\begin{center}

{\Large\bf  Quantum Corrections to Bekenstein--Hawking\\\vskip 0.4cm
Black Hole Entropy and Gravity Partition Functions }

\vspace{1cm}

{
{\large A. A. Bytsenko} $^{(a),}$
\footnote{aabyts@gmail.com}}
{\large and A. Tureanu} $^{(b),}$
\footnote{anca.tureanu@helsinki.fi}

\vspace{7mm}
$^{(a)}$ 
{\it
Departamento de F\'{\i}sica, Universidade Estadual de
Londrina, \\Caixa Postal 6001,
Londrina-Paran\'a, Brazil}

\vspace{0.1mm}
$^{(b)}$
{\it
Department of Physics, University of Helsinki,\\
P.O. Box 64, FI-00014 Helsinki, Finland}

\vspace{3mm}

\end{center}

\vspace{0.1in}
\begin{center}
{\bf Abstract}
\end{center}
Algebraic aspects of the computation of partition functions for quantum gravity and black
holes in $AdS_3$ are discussed. We compute the sub-leading quantum corrections to the 
Bekenstein--Hawking entropy. It is shown that the quantum corrections to the classical result 
can be included systematically by making use of the comparison with conformal field theory partition functions, via the $AdS_3/CFT_2$ correspondence. This leads to a better understanding of the role of modular and spectral functions, from the point of view of the representation theory of infinite-dimensional Lie algebras. Besides, the sum of known quantum contributions to the partition function can be presented in a closed form, involving the Patterson--Selberg spectral function. These contributions can be reproduced in a holomorphically factorized theory whose partition functions are associated with the formal characters of the Virasoro modules. We propose a spectral function formulation for quantum corrections to the elliptic genus from supergravity states.

\vfill




 \newpage
\tableofcontents

\renewcommand{\thefootnote}{\arabic{footnote}}
\setcounter{footnote}{0}

\section{Introduction}

In this paper we deal with applications of modular forms (and spectral functions related to the congruence subgroup of $SL(2, {\mathbb Z})$) to quantum gravity partition functions. The connection that can be established is particularly striking in the case of the correspondence between three-dimensional quantum gravity, in a space-time which is asymptotic 
to $AdS_{3}$, and the two-dimensional conformal field theory. We seek appropriate expressions for the partition functions and elliptic genera in general. Elliptic genera are natural topological invariants, and are also the one-loop string/gravity partition functions. They have been proven to be useful in the black hole entropy computations \cite{Strominger}.

Let us briefly recall the construction of the action of the Heisenberg/Clifford algebra on homology groups of varieties, which is key to our considerations. One can start with the  (integrable) highest-weight representation of the affine Lie algebra on the homology group of moduli spaces of torsion-free sheaves. The generators of the affine Lie algebra (as a Kac--Moody or Virasoro algebra) are given by moduli spaces of sheaves. Then the characters of highest-weight modules may be identified with the holomorphic parts of partition functions on the torus, for the corresponding field theories. This structures arises naturally, but not exclusively, in string theory and quantum gravity, and is particularly clear and treatable when supersymmetry is involved.

One of the purposes of the present work is to gain a better understanding of the role of the modular and spectral functions of hyperbolic geometry in the holomorphically factorized theory of three-dimensional gravity, from the point of view of the representation theory of infinite dimensional Lie algebras. Our particular interest in this example (as well as in an example of the elliptic genus from supergravity) stems from the $AdS_3/CFT_2$ correspondence. The geometric structure of three-dimensional gravity (and black holes) allows for exact computations, since its Euclidean counterpart is locally isomorphic to a constant curvature hyperbolic space. There is a correspondence between spectral functions related to Euclidean $AdS_3$ and modular-like functions (Macdonald series) \cite{Bo-By}. To be more precise, the classes of Euclidean $AdS_3$ spaces are quotients of the real hyperbolic space by a discrete group (a Schottky group). The boundaries of these spaces can be oriented compact surfaces with a conformal structure (compact complex algebraic curves).

We develop the space-time aspects of the quantum corrections for three-dimensional gravity and the elliptic genus from supergravity states. In these examples the appropriate Lie Virasoro algebra is a (universal) central extension of the Lie algebra of holomorphic vector fields on the punctured complex plane having finite Laurent series. It is for this reason that the Virasoro algebra plays a key role in conformal field theory. Besides the $AdS_3/CFT_2$ corespondence, we assume that the arguments of spectral functions of hyperbolic three-geometry take values on a Riemann surface, viewed as the conformal boudary of $AdS_3$. Thus the quantum correction can be rewritten in terms of the spectral functions of hyperbolic geometry, providing spectral flow (shift of the periodicities of quatum fields) and a kind of modular invariance.

The organization of the paper, and a brief summary of some of the results obtained, follows.
In Sect. 2 we analyze the holomorphic factorization for the one-loop correction to the 
three-dimensional gravity. We note that the holomorphic contribution to the
partition function corresponds to the formal character of the Virasoro module.
The symmetry group of $AdS_3$ gravity (with appropriate boundary conditions) is generated by the Virasoro algebra, and the one-loop partition function is indeed the partition function of a conformal field theory in two dimensions. We show that the infinite series of quantum corrections for the cases of three-dimensional gravity can be actually rewritten in terms of spectral functions in a holomorphically factorized theory.

There is a correspondence between quantum corrections to the black hole entropy and conformal field theory partition functions which can be reproduced, as is known, from gravity quantum corrections. This correspondence is made possible owing to the fact that the near horizon geometry of the black holes considered is locally $AdS_3$. This is most naturally viewed, because of that circumstance, as a beautiful example of the correspondence between three-dimensional anti-de Sitter gravity and two-dimensional conformal field theory. (A more simple geometrical structure of three-dimensional gravity and the associated black holes allows carrying out exact computations.) 

Owing to this correspondence, in Sect. 3 we analyze quantum corrections to the Bekenstein--Hawking black hole entropy \cite{Bekenstein, Hawking} and then in Sect. 4 we discuss sub-leading corrections. The asymptotic limit for the coefficient in the expansion of partition functions is calculated explicitly. The conclusion is that it has a universal form: it points out to sub-leading corrections, both to the entropy of the three-dimensional black hole, and also to the entropy of the conformal theory.

Special attention is devoted to the spectral function formulation of the one-loop quantum corrections in Sect. 5. We introduce the Patterson--Selberg and Ruelle spectral functions of hyperbolic three-geometry with an application to Macdonald polynomials.
From Sect. 5.3 on, we follow the strategy for analyzing the elliptic genus from supergravity, and turn to our interest, laying in supergravity contributions. This leads to a comparison with partition functions of CFT via the $AdS_3/CFT_2$ correspondence. 
Finally, in Sect. 6 we provide some conclusions.

\section{Holomorphic factorization of quantum corrections}
\label{Holomorphic}

One-loop corrections to three-dimensional gravity on $H^3/\Gamma$
are qualitatively similar to black hole quantum corrections. The
Euclidean black hole has an orbifold description
$H^3/\Gamma_{(a,b)}$ for suitable parameters $a>0$, $b\geq 0$,
where $H^{3}=\{(x,y,z)\in\mathbb{R}^3\mid z>0\}$ is the hyperbolic
three-space and $\Gamma_{(a,b)}\subset SL(2,\mathbb{C})$ is a
cyclic group of isometries. $H^3/\Gamma_{(a,b)}$ is a solution of
the Einstein equations $R_{ij}- (1/2)g_{ij}R_{g}-\Lambda_0
g_{ij}=0$ with negative cosmological constant $\Lambda_0$ (for
$\sigma=(-\Lambda_0)^{-1/2}$, the constant scalar curvature becomes
$R_{g}= 6\sigma^{-2}=-6\Lambda_0$). For our purpose, recall that the
(Chern--Simons) coupling constant is $k = \sigma/16G$ and, therefore,
$c= 24k \in {\mathbb Z}$. The dimensionless ratio $k$ is never a
variable parameter but does always take quantized values.
Actually, the fact that $k$ is not a continuous variable is
actually a more general consequence of the Zamolodchikov
$c$-theorem applied to the boundary of the conformal field theory.
We recall of the well-known fact that $24k= c_L=c_R= c$, $c$ being
the central charge of conformal field theory (which physically has
to do with the vacuum or Casimir energy). Then, we decompose
$q=\exp (2\pi i\tau) =\exp[2\pi(-{\rm Im}\tau +i{\rm Re}\tau)]$,
so that $|q\overline{q}|^{-k}=\exp (4\pi k {\rm Im}\tau)$, which is
the classical prefactor of the gravity partition function
\begin{equation}
{Z}_{\rm classical}(\tau, \overline{\tau}) = |q\overline{q}|^{-k}.
\label{class}
\end{equation}
For three-dimensional gravity in real hyperbolic space, the
one-loop contribution, as a product of holomorphic and
antiholomorphic functions, was analyzed in \cite{Maloney}:
\begin{eqnarray}
{Z}_{\rm gravity}^{\rm 1-loop}(\tau, \overline{\tau}) =  \prod_{n
=2}^{\infty}|1-q^n|^{-2} = \left[\prod_{n =2}^{\infty}(1-q^n)_{\rm
hol}\cdot \prod_{n=2}^{\infty}(1-\overline{q}^n)_{\rm antihol}
\right]^{-1}\,. \label{QGR}
\end{eqnarray}

A remarkable link between the theory of highest-weight modules
over the Virasoro algebra, conformal field theory and statistical
mechanics was discovered in \cite{Belavin1,Belavin2}. Here we
briefly note  some elements of the representation theory of the
Virasoro algebra which in fact are very similar to those for
Kac--Moody algebras. Let us consider the highest representation of
the Virasoro algebra. Let $M(c, h)$, with $c, h \in {\mathbb C}$, be
the Verma module over the Virasoro algebra (see, for example, 
\cite{Kac}). The {\it conformal central charge} $c$ acts on $M(c,
h)$ as $cI$. As $[{e}_0, {e}_{-j}] = n {e}_{-j}$, it follows that ${e}_0$ is
diagonalizable on $M(c, h)$, with the spectrum $h+ {\mathbb Z}_{+}$
and the eigenspace decomposition given by: $ M(c, h) =\bigoplus_{j\in
{\mathbb Z}_{+}} M(c, h)_{h+j}, $ where $M(c, h)_{h+j}$ is spanned
by elements of the basis $\{{e}_{-j_k}\}_{k=1}^n$ of $M(c, h)$.
The number $ {\rm dim}\, M(c, h)_{h+j} $ is the {\it classical
partition function}. This means that the Kostant partition
function (see Eq. (\ref{FK})) for the Virasoro algebra is the
classical partition function. On the other hand, the partition
functions can be rewritten in the form
\begin{equation}
{\rm Tr}_{M(c, h)}\, q^{{e}_0} := \sum_{\lambda}{\rm dim}\, M(c,
h)_{\lambda}\,q^{\lambda} = q^h\prod_{j\in {\mathbb Z}_+} (1-q^j)^{-1}\,.
\label{ch}
\end{equation}
The series ${\rm Tr}_V\,q^{{e}_0}$ is called the formal character of 
the Virasoro module $V$. 
We must emphasize the fact that the full quantum correction to the
gravity partition function admits the factorization (\ref{QGR}):
$Z_{\rm gravity}^{{\rm 1-loop}}(\tau, \overline{\tau}) =
Z(\tau)_{\rm hol}\cdot Z(\overline{\tau})_{\rm antihol}. $ Note that
the (anti)holomorphic contributions are similar to the formal character
of the Virasoro module (\ref{ch}).

\section{Quantum corrections to the black hole entropy}
\label{Entropy}

{\bf The three-dimensional black hole.} The corresponding metric
is, in spherical coordinates $(r, \varphi, \vartheta)$,
\begin{equation}
ds^2_{\rm Euclid} = (N_1(r)^2 + r^2N_2(r)^2)d\vartheta^2 +
N_1(r)^{-2}dr^2 +2r^2N_2(r)d\varphi d\vartheta + r^2d\varphi^2\,,
\end{equation}
where, for the mass and angular momentum parameters one has $M>
0$ and $J\geq 0$, respectively, while
\begin{equation}
N_1(r)^2= -M-\Lambda r^2 -J^2/4r^2\,,\,\,\,\,\,\,\,\,\,
N_2(r)=-J/2r^2\,.
\end{equation}
The periodicity of the Schwarzschild variable $\varphi$ means that
there is an identification $\varphi \sim \varphi + 2\pi n$, for
$n\in {\mathbb Z}$. It is quite remarkable that, for a suitable
change of variables $(r,\varphi, \vartheta)\rightarrow (x, y, z),
z>0$, the metric $ds^2_{\rm Euclid}$ transforms into the standard
hyperbolic metric $ds^2 = \sigma^2z^{-2}(dx^2+dy^2+dz^2)$ on
$H^3$. The metric $ds^2_{\rm Euclid}$ is a black hole solution
with outer and inner event radii $r_+$ and $ r_-$, respectively.
For $r_{+}>0, r_{-}\in i\mathbb{R}$ ($i^2=-1$, $r_{-}$ is purely
imaginary, since we are working with the Euclidean version), the
outer and inner horizons are respectively given by
\begin{equation}
r_{+}^2=\frac{M\sigma^2}{2}\left[1+\left(1+\frac{J^2}{M^2\sigma^2}\right)^{1/2}\right]
\,,\,\,\,\,\,\,\,\,\,\,\,\,\, r_{-}=-\frac{\sigma Ji}{2r_{+}}\,.
\label{eq:rs}
\end{equation}
$\Gamma_{(a,b)}:= \{\gamma^n_{(a,b)}\mid n\in\mathbb{Z}\}$ is
defined to be the cyclic subgroup of $SL(2,\mathbb{C})$, with
generator (see Eq. (\ref{group})) $ \gamma_{(a,b)} := {\rm
diag}(e^{a+ib}, \, \, \, e^{-(a+ib)}) $. A fundamental domain
$F_{(a,b)}$ for the action of $\Gamma_{(a.b)}$ on $H^3$ is given
by $ F_{(a,b)}=\{(x,y,z)\in H^3 \mid 1<x^2+y^2+z^2<e^{2a}\}. $ It
follows that $\Gamma_{(a,b)}$ is a {\it Kleinian} group.

{\bf The entropy.} The formula ${Z}_{\rm gravity}(\tau,
\overline{\tau}) = |q\overline{q}|^{-2k}{Z}_{\rm gravity}^{\rm
1-loop} (\tau, \overline{\tau})$ has a natural physical
interpretation. Indeed, it has the form of a trace,
\begin{equation}
Z ={\rm Tr}_{\cH(X)} q^{L_{0}} {\bar q}^{{\bar L}_{0}},
\end{equation}
over an irreducible representation of the Virasoro algebra;
$\cH(X)$ is the Hilbert space of the gravitation theory with the
target space $X= H^3/\Gamma$. The representation contains the ground
state $|0\rangle$ of weight $L_{0}|0\rangle = -k|0\rangle$, along
with its Virasoro descendants $L_{-n_{1}}\dots L_{-n_{i}}
|0\rangle$. Thus, with appropriate boundary conditions, the symmetry
group relevant to $AdS_{3}$ gravity is generated by the Virasoro
algebra.

Going back to black hole physics, observe that for the Lorentzian
form of the metric, $ds^2_{\rm Lorentz}$, the correspondence radii
$r_{\pm}$ are solutions of the equation $N_1(r)=0$, and have the
form
\begin{equation}
r_{\pm} = 4GM\sigma^2 \left[1 \pm \sqrt{1-
\left(J/M\sigma\right)^2}\right]\,.
\end{equation}
We assume that $1-(J/M\sigma)^2\geq 0$, i.e. $|J|\leq M\sigma$,
and $r_{\pm}\geq 0$. For the eigenvalues of the holomorphic and
antiholomorphic Virasoro operators $L_ 0$ and ${\overline L}_ 0$,
respectively, we have
\begin{eqnarray}
L_ 0 & = & (M\sigma + J)/2 = (r_+ + r_-)^2/16\sigma G, \label{L0}
\\
{\overline L}_ 0 & = & (M\sigma - J)/2 = (r_+ - r_-)^2/16\sigma G.
\label{L01}
\end{eqnarray}
The classical Bekenstein--Hawking entropy \cite{Bekenstein,Hawking} is $S_{\rm BH}\equiv {\rm
log} Z_{\rm classical} (\tau, \overline{\tau})$, where $Z_{\rm
classical} (\tau, \overline{\tau})$ is given by Eq.~(\ref{class}).
In fact, the partition function (\ref{QGR}) is a canonical ensemble
partition function of thermal $AdS_3$, so for $\beta \equiv {\rm
Im}\,\tau$, one obtains
\begin{equation}
S(\beta,|r_{-}|)  = {\rm log}\,{Z}_{\rm BH} - \beta Z_{\rm
BH}^{-1} \,\frac{\partial \,{Z_{\rm BH}}}{\partial \beta}\,.
\label{S}
\end{equation}

\section{Sub-leading corrections} \label{Sub-leading}

{\bf Digression: finite-dimensional Lie algebras.} Let ${\mathbb
R}^n$ be the $n$-dimensional real Euclidean space with the
standard basis $\varepsilon_1, \cdots , \varepsilon_n$ and the
bilinear form $(\varepsilon_i , \varepsilon_j)= \delta_{ij}$. All
lattices below will be sublattices of ${\mathbb R}^n$ with the
inherited bilinear form $(\cdot\mid\cdot)$. All indices are to be
distinct. Recall that a complex $n\times n$ matrix $A =
(a_{ij})_{i, j = 1}^n$ of rank $n$ is called a {\it generalized
Cartan matrix} if it satisfies the following conditions:

{\bf (i)} $a_{ii} = 2$ for $i = 1, \cdots, n$;

{\bf (ii)} $a_{ij}$ non-positive integers for $i \neq j$;

{\bf (iii)} $a_{ij} = 0$ implies $a_{ji} = 0$.

A realization of $A$ is a triple $\{{\mathfrak h}, \Pi, \Pi^\vee
\}$, where $\mathfrak h$ is a complex vector space, $\Pi =
\{\alpha_1, \cdots, \alpha_n\}\subset {\mathfrak h}^*$ and
$\Pi^\vee = \{\alpha_1^\vee, \cdots, \alpha_n^\vee\}$ are indexed
subsets in ${\mathfrak h}^*$ and $\mathfrak h$, respectively. We
also set $Q = \sum_{j=1}^n {\mathbb Z}\alpha_j$, \, $Q_+ =
\sum_{j=1}^n {\mathbb Z}_+\alpha_j$; the lattice $Q$ is called the
{\it root lattice}. Let us introduce the following root space
decomposition with respect to $\mathfrak h$: ${\mathfrak g}(A) =
\bigoplus_{\alpha\in Q}{\mathfrak g}_\alpha$, where ${\mathfrak
g}_\alpha = \{x\in {\mathfrak g}(A)\mid [h, x] = \alpha(h)x\,\,\,
\forall h\in {\mathfrak h}\}$ is the root space attached to
$\alpha$. In addition ${\mathfrak g}_0 = {\mathfrak h}$, the
number ${\rm mult}\, \alpha := {\rm dim}\, {\mathfrak g}_\alpha$ is
called the {\it multiplicity} of $\alpha$. An element $\alpha \in
Q$ is called a {\it root} if $\alpha \neq 0$ and ${\rm mult}\,
\alpha \neq 0$; a root $\alpha > 0$ (resp. $\alpha < 0$) is called
{\it positive} (resp. {\it negative}). Denote by $\triangle,
\triangle^+, \triangle^-$ the sets of all roots, positive and
negative roots respectively, such that $\triangle = \triangle^+\cup
\triangle^-$ (a disjoint union).

Let ${\mathfrak n}^+$ (resp. ${\mathfrak n} ^-$) denote the
subalgebra of ${\mathfrak g}(A)$ generated by $e_1, \cdots, e_n$
(resp. $f_1, \cdots, f_n$). Then we have the triangular
decomposition: ${\mathfrak g}(A) = {\mathfrak n}^-\oplus
{\mathfrak h}\oplus {\mathfrak n}^+$ (direct sum of vector
spaces). ${\mathfrak g}_\alpha\subset{\mathfrak n}^+$ if $\alpha >
0$ and ${\mathfrak g}_\alpha\subset{\mathfrak n}^-$ if $\alpha <
0$. It means that for $\alpha > 0$ (resp. $\alpha < 0$),
${\mathfrak g}_\alpha$ is the linear span of the elements of the
form $[\cdots [[e_{i_1}, e_{i_2}], e_{i_3}]\cdots e_{i_s}]$ (resp.
$[\cdots [[f_{i_1}, f_{i_2}], f_{i_3}]\cdots f_{i_s}]$) such that
$\alpha_{i_1}+ \cdots + \alpha_{i_s} = \alpha$ (resp. $=
-\alpha$). Besides ${\mathfrak g}_{\alpha_i} = {\mathbb C}e_i$,
${\mathfrak g}_{-\alpha_i} = {\mathbb C}f_i$, ${\mathfrak
g}_{s\alpha_i} = 0$ if $|s| > 1$. The {\it Chevalley involution}
of the Lie algebra ${\mathfrak g}(A)$ is determined by $w(e_i) = -
f_i,\, w(f_i) = -e_i,\, w(h) = -h$ if $h\in {\mathfrak h}$. Let
$\epsilon(w) \equiv {\rm det}_{{\mathfrak h}^*}w =
(-1)^{\ell(w)}$, where $\ell(w)$ is the length of $w$; also
$w({\mathfrak g}_\alpha) = {\mathfrak g}_{-\alpha}$, ${\rm mult}\,
\alpha = {\rm mult}\, (-\alpha)$ and \, $\triangle^- = -
\triangle^+$. Consider the expression
\begin{equation}
\prod_{\alpha \in \triangle^+}(1-e(-\alpha))^{- {\rm mult}\, \alpha}
= \sum_{\xi \in {\mathfrak h}^*} {\cK}(\xi) e(\xi)\,, \label{FK}
\end{equation}
defining a function $\cK$ on ${\mathfrak h}^*$ called the
(generalized) partition function (the symbol $\cK$ is in honour of Kostant). Note
that $\cK(\xi) = 0$, unless $\xi \in Q_+$; furthermore, $\cK(0) =
1$, and $\cK(\xi)$ for $\xi \in Q_+$ is the number of partitions
of $\xi$ into a sum of positive roots, where each root is counted
with its multiplicity. The last remark follows from another form
of formula  (\ref{FK}): $ \sum_{\xi \in Q_+} \cK(\xi) e(\xi) =
\Pi_{\alpha \in \triangle^+} (1 + e(\alpha) + e(2\alpha) + \cdots
)^{{\rm mult}\, \alpha}. $

Define $\triangle_0 = \{\alpha\in\triangle\mid{\overline \alpha}
=0\}$. The subgroup $W$ of $GL({\mathfrak h}^*)$ generated by all
fundamental reflections is called the Weyl group of ${\mathfrak
g}(A)$. The action of $r_i$ on ${\mathfrak h}^*$ induces the dual
fundamental reflections $r_i^\vee$ on $\mathfrak h$ (for the dual
algebra ${\mathfrak g}({}^tA)$). For each $i=1, \cdots, n$ we
define the fundamental reflection $r_i$ of the space ${\mathfrak
h}^*$ by $r_i(\lambda) = \lambda -\langle\lambda,
\alpha_i^\vee\rangle \alpha_i, \, \lambda\in {\mathfrak h}^*$. It
is clear that $r_i$ is a reflection since its fixed point set is
$T_i = \{\lambda\in {\mathfrak h}^*\mid\langle\lambda,
\alpha_i^\vee\rangle = 0\}$, and $r_i(\alpha_i) = -\alpha_i$. Let
$W_0$ be a (finite) subgroup of $W$ generated by reflections the
$r_\alpha$, with $\alpha\in \triangle_0$. We can describe the integrable
highest-weight modules $L(\Lambda)$ with respect to the algebras
${\mathfrak g}(A)$, where $A$ is a finite type matrix, and use the
specialization formula \cite{Kac}:
\begin{equation}
\prod_{\alpha\in \triangle^+\backslash \triangle_0}
(1-e(-{\overline \alpha}))^{{\rm mult}\,\alpha} = \sum_{w\in
W\backslash W_0}\epsilon(w)\cK(w(\rho))e(\overline{w(\rho)} -
\overline{\rho})\,. \label{Spec}
\end{equation}

{\bf Highest-weight modules: hyperbolic spaces.} Let ${\mathfrak
a}_0, {\mathfrak n}_0$ denote the Lie algebras of $A, N$ in an
Iwasawa decomposition, $G=KAN$. Since we are interested in
hyperbolic geometry, let us consider the case $G=SO_1(2n,1)$,
$K=SO(2n)$. The complexified Lie algebra ${\mathfrak g}={\mathfrak
g}^{\mathbb C}_0 ={\mathfrak so}(2n+1,{\mathbb C})$ of $G$ is of
the Cartan type $B_n$:
\begin{eqnarray}
B_n: & Q & =  \Big\{\sum_ik_i\varepsilon_i \in {\mathbb R}^n \mid
k_i\in {\mathbb Z}\Big\}\,,
\nonumber \\
& Q^\vee & = \Big\{\sum_ik_i\varepsilon_i\in {\mathbb
R}^n\mid k_i\in {\mathbb Z},\, \sum_ik_i\in 2{\mathbb Z}\Big\},
\nonumber \\
& \triangle & =  \{\pm\varepsilon_i\pm\varepsilon_j,
\pm\varepsilon_i\},
\nonumber \\
& \Pi & =  \{\alpha_1 = \varepsilon_1-\varepsilon_2, \cdots ,
\alpha_{n-1} = \varepsilon_{n-1} -\varepsilon_n, \alpha_n =
\varepsilon_n\},
\nonumber \\
& W & = \{{\rm all}\,\,\,\,{\rm permutations}\,\,\,\,{\rm
and}\,\,\,\,{\rm sign}\,\,\,\,{\rm changes} \,\,\,\,{\rm
of}\,\,\,\,{\rm the}\,\,\,\, \varepsilon_i\} = Aut\,Q\,.
\end{eqnarray}
Here, $\Pi$ is a basis of $Q$ over $\mathbb Z$ (the matrix
$(2(\alpha_i,\alpha_j)/(\alpha_i, \alpha_i))$ is the Cartan matrix
of the corresponding type). Since the rank of $G$ is one, $\dim
{\mathfrak a}_0=1$ by definition, say
\begin{equation}
{\mathfrak a}_0={\mathbb R}\,H_0 \,\,\,\, {\rm for}\,\,\,\,{\rm
a}\,\,\,\, {\rm suitable} \,\,\,\,{\rm basis}\,\,\,\, {\rm
vector}\,\,\,\, H_0 := {\rm antidiag}(1, \cdots, 1)
\end{equation}
is a $(n+1)\times(n+1)$ matrix. By this choice we have the
normalization $\beta(H_0)=1$, where $\beta: {\mathfrak
a}_0\rightarrow{\mathbb R}$ is the positive root which defines
${\mathfrak n}_0$. Note that the Killing form $(\ ,\ )$ is given by 
$(x,y)=(n-1)\,{\rm trace}(xy)$ for $x,\;y\in {\mathfrak g}_0$. 
The standard systems of positive roots $\triangle^{+},\triangle^{+}_s$ 
for ${\mathfrak g}$ and ${\mathfrak k}={\mathfrak k}^ {\mathbb C}_0$ -- the
complexified Lie algebra of $K$, with respect to a Cartan subgroup
$H$ of $G$,\, $H\subset K$, are given by
\begin{equation}
\triangle^{+}=\{\varepsilon_i|1\leq i\leq n\}\cup
\triangle^{+}_s\,, \,\,\,\,\,\,\,
\triangle^{+}_s=\{\varepsilon_i\pm \varepsilon_j|1\leq i<j\leq
n\},
\end{equation}
and
\begin{equation}
\triangle^{+}_n \stackrel{\rm def}{=}\{\varepsilon_i|1\leq i\leq
n\}
\end{equation}
is the set of positive non-compact roots. Here,
\begin{equation}
(\varepsilon_i , \varepsilon_j) = \frac{\delta_{ij}}{(H_0,
H_0)} = \frac{\delta_{ij}}{2(2n-1)}\,,\,\,\,\,\,\,\,
(\varepsilon_i \pm \varepsilon_j , \varepsilon_i \pm
\varepsilon_j) = \frac{1}{2n+1}\,\,\,,\,\,\, i<j\,,
\end{equation}
i.e. $ (\alpha , \alpha) = (2n-1)^{-1},\,\, \forall \alpha\in
\triangle^{+}_n.\, $ Let $\tau = \tau^{(j)}$ be a representation of
$K$ on $\Lambda^j{\mathbb C}^{2n}$. The highest weight of $\tau$,
$\Lambda_{\tau^{(j)}}=\Lambda_j$, is
\begin{equation}
\left\{ \begin{array}{ll} \varepsilon_1 + \cdots +
\varepsilon_j,\,\,\,\,\,\,
&{\rm if}\,\,\, j\leq n,\\
\varepsilon_1 + \cdots + \varepsilon_{2n-j},\,\,\,\,\,\, &{\rm
if}\,\,\, j>n.
\end{array} \right.
\end{equation}
Writing $ (\Lambda_j, \Lambda_j+2\rho_n) = (\Lambda_j, \Lambda_j)
+ (\Lambda_j, 2\rho_n), \,\, \rho_n = \sum_{i=1}^n
(n-i)\varepsilon_i $, for $j\leq n$ we have
\begin{eqnarray}
(\Lambda_j , \Lambda_j) & = & \left(\sum_{p=1}^j\varepsilon_p ,
\sum_{q=1}^j\varepsilon_q\right) = \sum_{p, q =1}^j (\varepsilon_p ,
\varepsilon_q) =\sum_{p=1}^j (\varepsilon_p , \varepsilon_p) =
\frac{j}{(H_0 , H_0)}\,,
\\
(\Lambda_j ,  2\rho_n) & = & \left(\sum_{p=1}^j\varepsilon_p ,
2\sum_{i=1}^j(n-i)\varepsilon_i +
2\sum_{i=j+1}^n(n-i)\varepsilon_i\right) = 2\sum_{p=1}^j(\varepsilon_p ,
(n-p)\varepsilon_p)
\nonumber \\
& = & \frac{2nj}{(H_0 , H_0)} -2\sum_{p=1}^jp(\varepsilon_p ,
\varepsilon_p) = \frac{2nj}{(H_0 , H_0)} -\frac{j(j+1)}{(H_0 ,
H_0)^2}.
\end{eqnarray}
Therefore,
\begin{equation}
(\Lambda_j , \Lambda_j+2\rho_n) = \frac{j(2n+1)}{(H_0 , H_0)} -
\frac{j(j+1)}{(H_0 , H_0)^2}.
\end{equation}
In the case $j>n$, we have
\begin{eqnarray}
(\Lambda_j , \Lambda_j) & = & \left(\sum_{p=1}^{2n-j}\varepsilon_p ,
\sum_{q=1}^{2n-j}\varepsilon_q\right) = \sum_{p=1}^{2n-j}(\varepsilon_p
, \varepsilon_p) =\frac{2n-j}{(H_0 , H_0)},
\\
(\Lambda_j , 2\rho_n) & = & 2\left(\sum_{p=1}^{2n-j}\varepsilon_p ,
\sum_{i=1}^{n}(n-i)\varepsilon_i\right)
\nonumber \\
& = & 2\left(\sum_{p=1}^{2n-j}\varepsilon_p ,
\sum_{i=1}^{2n-j}(n-i)\varepsilon_i +
\sum_{i=2n-j+1}^{n}(n-i)\varepsilon_i\right)
\nonumber \\
& = & 2\left(\sum_{p=1}^{2n-j}\varepsilon_p ,
\sum_{i=1}^{2n-j}n\varepsilon_i - \sum_{i=1}^{2n-j}i\varepsilon_i\right)
=\frac{2n(2n-j)}{(H_0 , H_0)} - 2\sum_{i=1}^{2n-j}i(\varepsilon_i
, \varepsilon_i)
\nonumber \\
& = & \frac{2n(2n-j) - (2n-j)(2n-j+1)}{(H_0 , H_0)} =
\frac{(2n-j)(j-1)}{(H_0 , H_0)}.
\end{eqnarray}
Thus, for $\Lambda_j = \triangle_{s}^{+}-$ highest weight of
$K=SO(2n)$ on $\Lambda^j{\mathbb C}^{2n}$, we have
\begin{equation}
(\Lambda_j , \Lambda_j + 2\delta_n) = \frac{2nj-j^2}{(H_0 , H_0)}
= \frac{2nj-j^2}{2(2n-1)} \,\,\,\,\,\,\, {\rm for}\,\,\,\, 0\leq
j\leq 2n.
\end{equation}
Let ${\mathfrak h}_0$ be the Lie algebra of $H$ and let
${\mathfrak h}^{*}_{\mathbb R}={\rm Hom}(\sqrt{-1}{\mathfrak h}_0,
{\mathbb R})$ be the dual space of the real vector space
$\sqrt{-1}{\mathfrak h}_0$. Thus, the $\{\varepsilon_i\}_{i=1}$
are an ${\mathbb R}$-basis of ${\mathfrak h}^{*}_{\mathbb R}$. Of
interest are the {\em integral} elements $f$ of ${\mathfrak
h}^{*}_{\mathbb R}$:
\begin{eqnarray}
f\stackrel{def}{=}\{\lambda\in {\mathfrak h}^{*}_{\mathbb R}\mid
\langle \lambda\mid\alpha\rangle\equiv \frac{2(\lambda,\alpha)}
{(\alpha,\alpha)}\in {\mathbb Z},\,\,\, \forall \alpha\in
\triangle^{+}\}.
\end{eqnarray}
Then we have
\begin{equation}
\langle \lambda\mid\varepsilon_i\rangle =  2\lambda_i\,\,\,\,\,
{\rm for}\,\,\,\,\, 1\leq i\leq n, \,\,\,\,\,\,\, \langle
\lambda\mid\varepsilon_i\pm\varepsilon_j\rangle
= \lambda_i\pm \lambda_j \,\,\,\,\,\,  {\rm for}\,\,\,\,\,1\leq
i<j\leq n,
\end{equation}
where we shall write $\lambda =\sum_{j=1}^n\lambda_j\varepsilon_j$
for $\lambda\in {\mathfrak h}^{*}_{\mathbb R},\, \lambda_j\in
{\mathbb R}$. Then clearly
\begin{eqnarray}
f & = & \{\lambda\in {\mathfrak h}^{*}_{\mathbb R}\mid
2\lambda_i\,\in \,{\mathbb Z}\}\,\,\,\,\,\,\,\,\,\,\,\,\,\, {\rm
for}\,\,\,\, 1\leq i\leq n,
\\
f & = & \{\lambda\in {\mathfrak h}^{*}_{\mathbb R}\mid
\lambda_i\pm \lambda_j \in \,{\mathbb Z}\}\,\,\,\,\,{\rm
for}\,\,\,\,1\leq i<j\leq n.
\end{eqnarray}
Let $ \rho_s= (1/2)\sum_{\alpha\in\triangle^{+}_s}\alpha,\,
\rho_n= (1/2)\sum_{\alpha\in\triangle^{+}_n}\alpha,\, \rho
=\rho_s+\rho_n= (1/2)\sum_{\alpha\in\triangle^{+}}\alpha. $ Then
\begin{equation}
\rho_s=\sum_{i=1}^n(n-i)\varepsilon_i,\,\, \rho_n=
\frac{1}{2}\sum_{i=1}^n\varepsilon_i,\,\, \rho =\sum_{i=1}^n(n-i-
\frac{1}{2})\varepsilon_i
\end{equation}
are all integral. The elements $\lambda$ of $f$ correspond to
characters $e^{\lambda}$ of $H$. We can deduce the specialization
formula \cite{Kac}:
\begin{equation}
\prod_{j\geq 1}(1-q^j)^{{\rm dim}\, {\mathfrak g}_j(s)} = \sum_{w\in
W^s}\varepsilon (w)\cK_s(w(\rho))q^{\langle\rho - w(\rho),
h^s\rangle}\,.
\end{equation}
Here,
\begin{equation}
\cK_s(\lambda) = \prod_{\alpha\in \triangle_{s}^{+}} \langle
\lambda, \alpha^{\vee}\rangle/\langle \rho_s,
\alpha^{\vee}\rangle,\,\,\,\,\,\,\, \triangle_{s}^{+} =
\{\alpha\in \triangle_+ \vert \langle\alpha, h^s\rangle = 0 \}
\end{equation}
and $\rho_s$ is the half-sum of roots from $\triangle_{s+}$; $W^s$
is a system of representatives of left cosets of the subgroup
$W_s$ generated by $r_\alpha$, $\alpha \in \triangle_{s+}$ in $W$,
so that $W= W_sW^s$; ${\mathfrak g}(A) = \oplus_j {\mathfrak
g}_j(s)$ is the $\mathbb Z$-gradation of ${\mathfrak g}(A)$ of
type $s$. For the case $G= SO_1(2n, 1), K=SO(2n)$ we have
\begin{equation}
\cK_s(\lambda) = \prod_{\alpha\in \triangle_{s}^{+}}(\lambda,
\alpha)/(\rho_s, \alpha) = \prod_{1\leq i<j\leq n}
\frac{\lambda_i^2 -\lambda_j^2}{(2n-i-j)(j-i)}\,.
\end{equation}

{\bf Asymptotic limit in the expansion of $\cK(N)$}. In general,
generating functions adopt the form of expressions for the
Poincar\'{e} polynomials $ \Pi_n(1-q^n)^{{\rm dim}\,{\mathfrak
g}_n}, $ $ \Pi_n(1-q^n)^{{\rm rank}\,{\mathfrak g}_n}\,. $ These
formulas are associated with the dimensions of the homology of
appropriate topological spaces (as the Euler--Poincar\'{e} identity
\cite{Fuks, Bo-By}). Formally, the sub-leading corrections to the
entropy $S$ can be associated with the Euler characteristic of an
appropriate complex or the coefficient in the product expansion:
\begin{eqnarray}
\prod_n(1-q^n)^{{\rm dim}\,{\mathfrak g}_{n}} & = &
\sum_{m,\lambda}(-1)^m q^\lambda {\rm
dim}\,H_m^{(\lambda)}({\mathfrak g})\,\, = \,\,
\sum_{\lambda}q^\lambda {\mathcal K}^{(\lambda)}({\mathfrak g}),
\\
{\mathcal K}^{(\lambda)}({\mathfrak g})  & = & \sum_{m}(-1)^m {\rm
dim}\,H_m^{(\lambda)}({\mathfrak g}),\,\,\,\,\,\,\, S
\Longrightarrow {\rm log}\,{\mathcal K}^{(\lambda)}({\mathfrak
g})\,.
\end{eqnarray}
We let 
\begin{equation}
\cQ(q) = \sum_{n\geq 0}q^n \cK(n) = \prod_{n\in {\mathbb Z}_+}(1-q^n)^{-1},
\end{equation}
then \cite{Andrews}
\begin{equation}
\cQ(\exp(2\pi i(h + iz)/k)) = \Omega_{h,k}z^{1/2}\exp (\pi(z^{-1}
- z)/12k)\cQ\exp((2\pi i(h' + iz^{-1})/k)),
\end{equation}
where ${\rm Re}\, z> 0$, the principal branch of $z^{1/2}$ is
selected, $h'$ is a solution of the congruence $h h' = - 1\, ({\rm
mod}\, k)$, and $\omega_{h, k}$ is a 24 k-th root of unity given
by
\begin{equation}
\Omega_{h, k} = \left\{ \begin{array}{ll} (\frac{-k}{h})\exp\{-\pi
i
(\frac{1}{4}(2-hk-h)+\frac{1}{12}(k-k^{-1})(2h-h^{\prime}+h^2h^{\prime}))\},
\,\,\,\,\, &{\rm for}\,\,\,\, h\,\,\,\,{\rm odd},
\\
\\
(\frac{-h}{k})\exp\{-\pi i
(\frac{1}{4}(k-1)+\frac{1}{12}(k-k^{-1})(2h-h^{\prime}+h^2h^{\prime}))\},
\,\,\,\,\, &{\rm for}\,\,\,\, k\,\,\,\,{\rm even},
\end{array} \right.
\end{equation}
with $(a/b)$ the Legendre--Jacobi symbol. There is the elegant
representation of Rademacher for $\omega_{h, k}$: $\omega_{h, k} =
\exp\{\pi is(h, k)\}$, where $s(h, k)$ is the Dedekind sum:
$s(h,k) = \sum_{\mu=1}^{k-1}(\mu/k- [\mu/k]-1/2)(h\mu/k-
[h\mu/k]-1/2)$. Clearly, Cauchy's integral theorem implies that
\begin{eqnarray}
\cK(n) & = & \frac{1}{2\pi i}\int_{\cC}\frac{\cQ(q)}{q^{n+1}}dq =
\rho^{-n}\int_0^1 \cQ[\rho\exp\{2\pi i \varphi\}]\exp \{-2\pi i
n\varphi\}d\varphi
\nonumber \\
& = & \exp\left\{\frac{2\pi n}{N^2}\right\}\sum_{k=1,\, (h,k)=1,\,
0\leq h<k} \exp \left\{-\frac{2\pi i h n}{k}\right\}\Omega_{h, k}
\nonumber \\
& \times & \int_{-\theta^{\prime}_{h,
k}}^{\theta^{\prime\prime}_{h,k}}
z^{1/2}\exp\left\{\pi\frac{z^{-1}-z}{12k}\right\}
\cQ\left[\exp\left\{2\pi
i\frac{h^{\prime}+iz^{-1}}{k}\right\}\right] \exp\{-2\pi
in\varphi\}d\varphi.
\end{eqnarray}
Here, $\theta^{\prime}_{0, 1} = (N+1)^ {-1}, \theta^{\prime}_{h, k}
= h/k- (h_0+h)/(k_0+h), \theta^{\prime \prime}_{h, k} =
(h_1+h)/(k_1+h)-h/k$, and $h_0/k_0, h/k$, $h_1/k_1$ are the three
successive terms in the set of proper Farey fractions $F_N$ of
order $N$ (see for details \cite{Andrews}).

However the coefficient in the expansion of the generating
functionals in its final form is not always known. We will here
simplify the calculation and apply an asymptotic limit for that
coefficient. We will use the Meinardus theorem
\cite{Meinardus1,Meinardus2}; for the first time this theorem has
been used for the $p$-brane quantum states density in
\cite{PhysReports}. It gives a possibility to improve the Cardy
formula (including a prefactor). We set
\begin{equation}
\prod_{n\in {\mathbb Z}_+} \left(1- q^{n+\varepsilon}\right)^{-d_n} = 1 +
\sum_{N\in {\mathbb Z}_+}\cK(N)q^{N +\varepsilon}\,, \label{serie}
\end{equation}
where, as before, $\exp (2\pi i\tau) = \exp (-2\pi\,{\rm Im}\,\tau +
2\pi i {\rm Re}\,\tau)$,\, ${\rm Im}\,\tau >0$\, ($|q|< 1$),\,
$\varepsilon \geq 0$ and $d_n \equiv {\rm dim}\,{\mathfrak g}_n \,
({\rm or\,\, rank}\,{\mathfrak g}_n) > 0$. Let
\begin{equation}
{\mathfrak D}(s; \varepsilon)= \sum_{N\in {\mathbb Z}_+}d_N
(N+\varepsilon)^{-s} \,,\,\,\,\,\,\,\,\, s= \sigma +i\rho,
\end{equation}
be the associated Dirichlet series, which converges for
$0<\sigma<p$. Suppose that ${\mathfrak D}(s;\varepsilon)$ can be
analytically continued into the region $\sigma\geq -C_0\, (0< C_0
< 1)$ and that, here, ${\mathfrak D}(s; \varepsilon)$ is analytic,
except for a pole of order one at $s=p$ with residue $\cA$. We
also assume that ${\mathfrak D}(s;\varepsilon) =
O(|\rho|^{C_1})$ uniformly at $|\rho|\rightarrow \infty$,
where $C_1$ is a fixed positive real number. Expanding $\cK(N)$, one arrives at a complete asymptotic limit
\begin{eqnarray}
\cK(N)_{N\rightarrow \infty} & = & \cC(p)N^{\frac{2{\mathfrak
D}(0;\varepsilon)-p-2}{2(1+p)}} \exp\left\{\frac{1+p}{p}[\cA
\Gamma(1+p)\zeta_R(1+p)]^ {\frac{1}{1+p}}N^{\frac{p}{1+p}}\right\}
\nonumber \\
& \times & [1+{O}(N^{-{\kappa}})]\,, \label{B(N)}
\\
{\cC}(p) & = & [\cA \Gamma(1+p) \zeta_R (1+p)]^{\frac{1-2{\mathfrak
D}(0;\varepsilon)}{2p+2}} \cdot \frac{\exp\left[(d/ds){\mathfrak
D}(0;\varepsilon)\right]} {[2\pi(1+p)]^{1/2}}\,, \label{C(p)}
\end{eqnarray}
where $ {\kappa}  = p/(1+p)\cdot {\rm min} (C_0/p - \delta/4,
1/2-\delta), $ $ 0<\delta<2/3, $ and $\zeta_R(s)$ is the Riemann
zeta function. We should  stress the important physical
significance of the coefficients $\cK(N)$. The results
(\ref{B(N)}) and (\ref{C(p)}) have a universal character for the
generating functions associated with the $q$-series of the modular
forms. Finally,
\begin{eqnarray}
S(N) & := &  {\rm log}\, \cK (N) \approx \cC_1(p)
N^{\frac{p}{p+1}} + (\cC_2(p) + 2\kappa){\rm log}\,N\,,
\label{entropy-final}
\\
\cC_1(p)  & \equiv &   \frac{1+p}{p}[\cA
\Gamma(1+p)\zeta_R(1+p)]^{\frac{1}{1+p}}\,, \,\,\,\,\,\,\,\,\,\,
\cC_2(p) \equiv \frac{2{\mathfrak D}(0;\varepsilon)-p-2}{2(1+p)}\,.
\end{eqnarray}

{\bf The entropy of conformal field theory.} The holomorphic
contribution to the full gravity partition function becomes
\begin{eqnarray}
Z(\tau)_{\rm hol} =
q^{-k}\prod_{n\in {\mathbb Z}_+}(1-q^{n+1})^{-1}\,\, \equiv \,\,
q^{-k}(1 + \sum_{N\in {\mathbb Z}_+} \cK(N)q^{N +1})\,.
\label{holomorphic}
\end{eqnarray}
Comparing this expression with Eq. (\ref{serie}) we find: $d_n = 1,\,
\varepsilon = 1$, and therefore ${\mathfrak D}(s; 1)= \zeta_R(s; 1)
\equiv \zeta_R(s)$. In Eq. (\ref{entropy-final}) the second term
is the logarithmic correction. Typically, this term appears when
the entropy is computed in the microcanonical ensemble (as
opposite to the canonical one) \cite{Maloney}. The explicit value
of the prefactor $\cC(p)$ in the expansion (\ref{C(p)}) gives the
constant term in the final Eq.~(\ref{entropy-final}). Our goal is
to compute the entropy in a black hole geometry, where the
partition function ${\rm STr}_{\cH}q^{{\widehat D}}$ for an
appropriate operator insertion $\widehat D$ is calculated over the
brane Hilbert space $\cH$. Using Eq. (\ref{entropy-final}) we get
the final result: 
\begin{equation}
S_{\rm CFT}(N) \approx \pi \left[ \left(2\!\!\sum_{(j\,\,{\rm even})}h_j +
\sum_{(j\,\,{\rm odd})}h_j\right)/3\right]^{1/2}N^{1/2}  + (2\kappa - 1){\rm
log}\,N\,.
\end{equation}
We would like to comment that, improving on Cardy's saddle point approximation, it has been shown in \cite{SSJ} that for any unitary, modular invariant $2d$ CFT with discrete spectrum (i.e. when partition function is a holomorphic function of a power of $q$ and $\bar q$), one may capture not only $\log N$ corrections as in (47), but also all polynomially suppressed $1/N^k$ corrections to the entropy. There is in fact a closed form for the entropy as $\log (I_1(S_{Cardy})/S_{Cardy})$. Moreover, it has been shown that in this approximation and up to exponentially suppressed $\exp(-N)$ corrections, the partition of any such $2d$ CFT is holomorphically factorizable \cite{SSJ}.

\section{Space-time aspects of the computation of gravity partition functions}

\subsection{Spectral functions of hyperbolic three-geometry}

If $L_p$ is a self-adjoint Laplacian on $p$-forms, then the
following results hold. There exist $\varepsilon,\delta >0$ such
that for $0<t<\delta$ the heat kernel expansion for Laplace
operators on a compact manifold $X$ is given by $ {\rm
Tr}\left(e^{-tL_p}\right)= \sum_{0\leq \ell\leq \ell_0} a_\ell
(L_p)t^{-\ell}+ {O}(t^\varepsilon). $ The coefficients
$a_\ell(L_p)$ are called Hadamard--Minakshisundaram--De Witt--Seeley
coefficients (or, sometimes, heat kernel, or just heat
coefficients). Let $\chi$ be an orthogonal representation of
$\pi_1(X)$. Using the Hodge decomposition, the vector space
$H(X;\chi)$ of twisted cohomology classes can be embedded into
$\Omega(X;\chi)$ as the space of harmonic forms. This embedding
induces a norm $|\cdot|^{RS}$ on the determinant line ${\rm
det}\,H(X;\chi)$. The Ray--Singer norm $||\cdot||^{RS}$ on ${\rm
det}\,H(X;\chi)$ is defined by \cite{Ray}
\begin{equation}
||\cdot||^{RS}\stackrel{def}=|\cdot|^{RS}\prod_{p=0}^{{\rm
dim}\,X} \left[\exp\left(-\frac{d}{ds} \zeta
(s|L_p)|_{s=0}\right)\right]^{(-1)^pp/2} \mbox{,}
\end{equation}
where the zeta function $\zeta (s|L_p)$ of the Laplacian acting on
the space of $p$-forms orthogonal to the harmonic forms has been
used. For a closed connected orientable smooth manifold of odd
dimension and for the Euler structure $\eta\in {\rm Eul}(X)$, the
Ray--Singer norm of its cohomological torsion
$T_{an}(X;\eta)= T_{an}(X)\in {\rm det}H(X;\chi)$ is equal to
the positive square root of the absolute value of the monodromy of
$\chi$ along the characteristic class $c(\eta)\in H^1(X)$:
$||T_{an}(X)||^{RS}=|{\rm det}_{\chi}c(\eta)|^{1/2}$. In the
special case where the flat bundle $\chi$ is acyclic, we have
\begin{equation}
\left[T_{an}(X)\right]^2 =|{\rm det}_{\chi}c(\eta)|
\prod_{p=0}^{{\rm dim}\,X}\left[\exp\left(-\frac{d}{ds} \zeta
(s|L_p)|_{s=0}\right)\right]^{(-1)^{p+1}p}\,.
\label{RS}
\end{equation}

For a  closed oriented hyperbolic three-manifolds of the form $X =
{H}^3/\Gamma$, and for acyclic $\chi$, the $L^2$-analytic torsion
has the form \cite{Fried,Bytsenko3}: $[T_{an}(X)]^2={\mathcal
R}(0)$, where ${\mathcal R}(s)$ is the Ruelle function (it can be
continued meromorphically to the entire complex plane $\mathbb
C$). 

Recall that the Euclidean
sector of $AdS_3$ has an orbifold description $H^3/\Gamma$. The
complex unimodular group $G=SL(2, {\mathbb C})$ acts on the real
hyperbolic three-space $H^3$ in a standard way, namely for
$(x,y,z)\in H^3$ and $g\in G$, one has $g\cdot(x,y,z)= (u,v,w)\in
H^3$. Thus, for $r=x+iy$,\, $g= \left[
\begin{array}{cc} a & b \\ c & d \end{array} \right]$, $ u+iv =
[(ar+b)\overline{(cr+d)}+ a\overline{c}z^2]\cdot [|cr+d|^2 +
|c|^2z^2]^{-1},\, w = z\cdot[ {|cr+d|^2 + |c|^2z^2}]^{-1}\,, $
where the bar denotes complex conjugation. Let $\Gamma \in G$ be
the discrete group of $G$ defined as
\begin{eqnarray}
\Gamma & = & \{{\rm diag}(e^{2n\pi ({\rm Im}\,\tau + i{\rm
Re}\,\tau)},\,\,  e^{-2n\pi ({\rm Im}\,\tau + i{\rm Re}\,\tau)}):
n\in {\mathbb Z}\} = \{{\gamma}^n:\, n\in {\mathbb Z}\}\,,
\nonumber \\
{\gamma} & = & {\rm diag}(e^{2\pi ({\rm Im}\,\tau + i{\rm
Re}\,\tau)},\,\,  e^{-2\pi ({\rm Im}\,\tau + i{\rm Re}\,\tau)})\,.
\label{group}
\end{eqnarray}
One can construct a zeta function of Selberg-type for the group
$\Gamma \equiv \Gamma_{(a, b)}$ generated by a single hyperbolic
element of the form ${\gamma_{(a, b)}} = {\rm diag}(e^z, e^{-z})$,
where $z= a+ib$ for $a, b >0$. Actually we will take $a = 2\pi\,
{\rm Im}\,\tau$ and $b = 2\pi\, {\rm Re}\,\tau$. Then the
Patterson--Selberg spectral function $Z_\Gamma (s)$ which can be
attached to $H^3/\Gamma$ has the form:
\begin{equation}
Z_\Gamma(s) :=\prod_{\stackrel{k_1,k_2\geq
0}{k_1,k_2\in\mathbb{Z}}}[1-(e^{ib})^{k_1}(e^{-ib})^{k_2}e^{-(k_1+k_2+s)a}]\,.
\label{zeta00}
\end{equation}
The zeros of $Z_\Gamma (s)$ are precisely the complex numbers $
\zeta_{n,k_{1},k_{2}} = -\left(k_{1}+k_{2}\right)+i\left(k_{1}-
k_{2}\right)b/a + 2\pi i n/a,\, n \in {\mathbb Z}\,, $ and the
logarithm of $Z_\Gamma (s)$, for ${\rm Re}\, s> 0$, is given by
\cite{Bytsenko07}
\begin{equation}
{\rm log}\, Z_\Gamma (s)  = - \sum_{n\in {\mathbb Z}_+}\frac{e^{-n
a (s-1)}} {4n[\sinh^2\left(\frac{a n}{2}\right)
+\sin^2\left(\frac{b n}{2}\right)]}\,. \label{logZ}
\end{equation}
It can also be shown that the zeta function $Z_\Gamma (s)$ is an
entire function of order three and finite type. Let us introduce next the Ruelle function $\cR(s)$. The function ${\mathcal R}(s)$ is an alternating product of more complicate factors, each of which is a spectral function $Z_{\Gamma}(s)$.
Using expression (\ref{logZ}), we get (for details, see \cite{Bytsenko07,Bo-By,BBE})
\begin{eqnarray}
\prod_{n=\ell}^{\infty}(1- q^{\mu n+\varepsilon}) 
& = & \prod_{p=0, 1}Z_{\Gamma}(\underbrace{(\mu\ell+\varepsilon)(1-i\varrho(\tau)) 
+ 1 -a}_s + a(1 + i\varrho(\tau)p)^{(-1)^p} 
\nonumber \\
& = &
\cR(s = (\mu\ell + \varepsilon)(1-i\varrho(\tau)) + 1-a),
\\
\prod_{n=\ell}^{\infty}(1+ q^{\mu n+\varepsilon}) 
& = & 
\prod_{p=0, 1}Z_{\Gamma}(\underbrace{(\mu\ell+\varepsilon)(1-i\varrho(\tau)) + 1-a + 
i/(2\,{\rm Im}\,\tau)}_s
+ a(1+ i\varrho(\tau)p))^{(-1)^p}
\nonumber \\
& = & 
\cR(s = (\mu\ell + \varepsilon)(1-i\varrho(\tau)) + 1-a + i/(2\,{\rm Im}\,\tau))\,, 
\end{eqnarray}
with $q\equiv e^{2\pi i\tau}$, $\varrho(\tau) = {\rm Re}\,\tau/{\rm Im}\,\tau$, 
$\mu$ -- a real number, $\ell \in {\mathbb Z}_+$  and $\varepsilon \in {\mathbb C}$. 
We can use the Ruelle function $\cR(s)$ to write the results in a most general form.
Taking $\nu\in {\mathbb C}$, then
\begin{eqnarray}
\prod_{n=\ell}^{\infty}(1-q^{\mu n+ \epsilon})^{\nu n} & = & 
\cR(s=(\mu\ell + \varepsilon)(1-i\varrho(\tau))+1-a)^{\nu\ell}
\nonumber \\
& \times &
\!\!\!
\prod_{n=\ell+1}^{\infty} 
\cR(s=(\mu n + \varepsilon)(1-i\varrho(\tau))+1-a)^{\nu}\,,
\label{RU1}
\\
\prod_{n=\ell}^{\infty}(1+q^{\mu n+ \epsilon})^{\nu n} & = & 
\cR(s=(\mu\ell + \varepsilon)(1-i\varrho(\tau))+1-a+ i/(2\,{\rm Im}\,\tau))^{\nu\ell}
\nonumber \\
& \times &
\!\!\!
\prod_{n=\ell+1}^{\infty} 
\cR(s=(\mu n + \varepsilon)(1-i\varrho(\tau))+1-a+ i/(2\,{\rm Im}\,\tau))^{\nu}\,.
\label{RU2}
\end{eqnarray}

\subsection{Quantum corrections for three-dimensional gravity} 

Besides the $AdS_3/CFT_2$ correspondence, we assume that the arguments of spectral functions of hyperbolic three-geometry take values on a Riemann surface, viewed as the conformal boundary of $AdS_3$. Thus the quantum correction can be rewritten in terms of the spectral
functions as follows
\begin{equation}
{Z}_{\rm gravity}^{\rm 1-loop}(\tau, \overline{\tau})  = \prod_{n
=2}^{\infty}|1-q^n|^{-2} =  \left[ \cR(s= 2-2i\varrho(\tau))_{\rm
hol}\cdot \cR({s}=2+2i\varrho(\tau))_{\rm antihol}\right]^{-1}\,.
\end{equation}
It is known that one-loop corrections to three-dimensional gravity
in locally Anti-de Sitter space-times are qualitatively similar to
black hole quantum corrections. The simple geometrical structure
of three-dimensional gravity allows to perform exact computations,
since its Euclidean counterpart is locally isomorphic to a
constant curvature hyperbolic space, $H^3$. In the physical
literature it is usually assumed that the fundamental domain for
the action of a discrete group $\Gamma$ has finite volume. On the
other hand, a three-dimensional black hole has an Euclidean
quotient representation $H^3/\Gamma$, for an appropriate $\Gamma$,
where the fundamental domain has infinite hyperbolic volume (for
the non-spinning black hole, one can choose $\Gamma$ to be the
Abelian group generated by a single hyperbolic element
\cite{Perry}). For the discrete groups of isometries of
three-dimensional hyperbolic space with a fundamental domain of
infinite volume (e.g., for Kleinian groups), Selberg-type
functions and trace formulas (excluding fundamental domains with
cusps) have been considered in \cite{Perry1}. Note that things are
quite difficult in the case of an infinite-volume setting, due to
the infinite multiplicity of the continuous spectrum and to the
absence of a canonical renormalization of the scattering operator
to render it trace-class. However, for a three-dimensional black
hole one can bypass much of the general theory and try to proceed
more directly, by defining a Selberg function attached to
$H^3/\Gamma$ and establishing a trace formula, which is a version
of Poisson's summation formula for resonances (for details, see
\cite{Perry}). In fact, there is a special relation between the
spectrum and the {\it truncated} heat kernel of the Euclidean
black hole and the Patterson--Selberg spectral function
\cite{Bytsenko07}. From (\ref{S}) we get
\begin{equation}
S(\beta,|r_{-}|)  = (1- \beta \frac{\partial}{\partial \beta})
\,\,{\rm log}\left\{ {Z}_{\rm classical}\times\, [\cR(s=2+
2i|r_{-}|\beta^{-1})\, \cdot \cR({s}=2-
2i|r_{-}|\beta^{-1})]^{-1} \right\}\,. \label{S1}
\end{equation}

\subsection{Partition functions from supergravity}

In the case of supergravity, sub-leading corrections to the semi-classical result also can be included systematically, laying the groundwork for comparison with partition functions of conformal field theory via the $AdS_3/CFT_2$ correspondence \cite{Kraus,Gaiotto}. We follow 
the strategy for analyzing the elliptic genus from supergravity, and turn  to our interest in the supergravity contributions.

Consider the contribution from supergravity states, which can be obtained, for
example, from the fluctuation spectrum of supergravity
compactified on $AdS_3$ times some compact space $X$. Generalizing
our computation we make references to the two chiralities of the
conformal field theory with the convention that holomorphic and anti-holomorphic correspond to 
left and right. Besides the Virasoro algebras, we pay close
attention to $U(1)$ and R-symmetry current algebras.

Let us consider the $(0,4)$ case, corresponding to M-theory on
$AdS_3 \times S^2 \times X$. The $(0,4)$ conformal field theory on the $AdS_3$
boundary describes M5-branes wrapped on 4-cycles in $X_6$
\cite{Maldacena} (the same conformal field theory also describes black rings
\cite{Bena}.) To define the elliptic genus we introduce potentials
for the charges $q^I,\,{\overline q}^I$; the R-charge is $q^0$
(for details, see \cite{Kraus}). Up to a spectral flow, supergravity
states can carry vanishing charges, $q^I=0$.  These charges are
instead carried by wrapped branes. So the contribution to the
polar part of the elliptic genus from such supergravity states is
$ \chi^{{\rm sugra}}(\tau) = \sum_m \cC^{\rm sugra}(m) q^m, $
where $q= \exp(2\pi i \tau)$ (in order to extract the coefficients
$\cC^{\rm sugra}(m) $ we have to compute $\chi^{{\rm
sugra}}(\tau)$).

Let us take into account the NS sector; the elliptic genus in the
NS sector related to the R sector by so-called spectral flow
\cite{Kraus}. The contribution from supergravity states to the NS
sector elliptic genus  can be written $ \chi_{\rm NS}^{\rm
sugra}(\tau) = {\rm Tr}_{\cH_{\rm cp}} ((-1)^{\tilde{q}^0}
q^{L_0})\,, $ where the trace is over the space $\cH_{\rm cp}$ of
chiral primaries. The elliptic genus receives contributions from
rightmoving chiral primaries obeying $\tilde{h}= (1/2)\tilde{q}^0$
\cite{Kraus}, where $\tilde{h}$ is the eigenvalue of the Virasoro
operator $\tilde{L}_0$.

Consider the single particle spectrum; suppose it starts at
$h_{\rm min} = \tilde{h}_{\rm min}+s$ for some $s$ 
\footnote{ 
The appropriate complete spectrum of single particle primaries the
reader can find, for example, in \cite{Kraus}. Note that
multiparticle chiral primaries can be obtained by taking arbitrary
tensor products of single particle chiral primaries. 
}. 
Taking into account a bosonic contribution $
\Pi_{\ell, p =0}^\infty\sum_{m=0}^\infty q^{m(h_{\rm
min}+\ell+p)}= \Pi_{\ell, p =0}^\infty (1-q^{\tilde{h}_{\rm
min}+s+\ell+p})^{-1}, $ where $m$ stands for the number of
particles, $p$ for acting with $(L_{-1})^p$, and $\ell$ for
$\tilde{h} = \tilde{h}_{\rm min}+\ell$, one can define
$n=\ell+p+1$, and for bosons and fermions contributions we get (see Eq. (\ref{RU1})):
\begin{eqnarray}
\chi_{\rm NS}^{\rm boson}(\tau) & = & \prod_{n\in {\mathbb Z}_+} (1-
q^{h_{\rm min}-1+n})^{-n}  = \prod_{n\in {\mathbb Z}_+} [\cR(s=(n+h_{\rm
min}-1)(1-i\varrho(\tau)))]^{-1}\,, \label{boson}
\\
\chi_{\rm NS}^{\rm fermion}(\tau) & = & \prod_{n\in {\mathbb Z}_+} (1-
q^{h_{\rm min}-1+n})^n \,\,\,\,\,\, = \prod_{n\in {\mathbb Z}_+}
\cR(s=(n+h_{\rm min}-1)(1-i\varrho(\tau)))\,. \label{fermion}
\end{eqnarray}

{\bf Example: compactification on $AdS_3\times S^2\times CY_3$.}
In the case of a 5-dimensional supergravity obtained by
compactifying M-theory on $X = CY_3$, the 5-dimension (massless)
spectrum is written in the $\cN=2$ language in terms of the number
of: 
\begin{equation}
\left\{ \begin{array}{ll} 
{\rm vectormultiplets}\,\,\,\,\,\,\,\,\,\,\,\,\,\,\,  n_V = h^{1, 1}-1
\\
{\rm hypermultiplrets}\,\,\,\,\,\,\,\,\,\,\,\,\,\, n_H=
2(h^{2,1}(X)+1)
\\
{\rm gravitino}\,\,\,\,{\rm  multiplets}\,\,\,\, n_S,\,\,\,\, {\rm in}\,\,\,\,
{\rm addition}\,\,\,\, {\rm to}\,\,\,\, {\rm the}\,\,\,\,
{\rm graviton}\,\,\,\, {\rm multiplet}
\end{array} \right.
\end{equation}
In addition, $h^{i, j}$ are generators of
degree $(i, j)$ \cite{Boer}. The spectrum on $AdS_3\times S^2$ organizes into representations of $SL(2, {\mathbb R})\times SU(1, 1\mid 2)$.  
\footnote{ Recall that the Lie superalgebra
$SU(1, 1\mid 2)$ is defined by the super-commutation relations
among the fourteen generators $Y_{\mu}$ $(\mu = 1,2,3,\cdots,
14)$, $[Y_\mu\,,\,Y_\nu] = Y_\mu Y_\nu - (-1)^{p(\mu)p(\nu)}Y_\nu
Y_\mu = if_{\mu\nu\rho} Y_\rho$. Here $f_{\mu\nu\rho}$ are
structure constants, the fermion number $p(\mu)$ is 0 if $\mu \in
\{1, 2, \cdots, 6\}$, or 1 if $\mu \in \{7, 8, \cdots, 14\}$.} 
The superconformal relations for the finite subalgebra  $\{L_0, L_{\pm 1}, T_0^i, G_{\pm 1/2}^i, \overline{G}_{\pm 1/2}^i\}$ of
the minimal $\cN=4$ superconformal algebra \cite{Fujii} in the NS sector are: 
\begin{eqnarray}
[L_{m},L_{n}] & = & (m-n)L_{m+n}+ km((m^{2}-1)/2)\,\delta_{m+n,0},
\nonumber \\
\{ G^{a}_{r},G^{b}_{s}\} & = & \{ {\overline G}^{a}_{r},{\overline
G}^{b}_{s}\}=0,
\nonumber \\
\{ G^{a}_{r}, {\overline G}^{b}_{s}\} & = &
2\delta^{ab}L_{r+s}-2(r-s)\sigma^{i}_{ab}T^{i}_{r+s} +
k((4r^{2}-1)/2)\,\delta_{r+s,0},
\nonumber \\
\left[T^{i}_{m}, T^{j}_{n}\right] & = &
i\varepsilon^{ijk}T^{k}_{m+n} + (km/2)\,
\delta_{m+n,0}\delta^{ij},
\nonumber \\
\left[T^{i}_{m}, G^{a}_{r}\right] & = & -
(1/2)\,\sigma^{i}_{ab}G^{b}_{m+r}, \quad [T^{i}_{m},{\overline
G}^{a}_{r}]= - (1/2)\,\sigma^{i*}_{ab}{\overline G}^{b}_{m+r},
\nonumber \\
\left[L_m,G_r^a\right] & = & (m/2 -r)\, G_{m+r}^a, \quad
[L_m,{\overline G}^{a}_{r}]= (m/2 -r) \,{\overline G}^{a}_{m+r},
\nonumber \\
\left[L_m,T^{i}_{n}\right] & = & -nT^{i}_{m+n}\,, \nonumber
\end{eqnarray}
In these formulas $\sigma^i$ is the Pauli spin matrix, $m, n$ run
over integers, $r, s$ are half odd integers, $a, b$ = 1 or 2,
while $i$ is the $SU(2)$ index taking the values 1, 2 or 3. A state
$|\phi\rangle$ is said to be chiral primary if
$G^2_{-1/2}|\phi\rangle = \overline{G}^1_{-1/2}|\phi\rangle =0$,
$G^a_{n+1/2}|\phi\rangle = \overline{G}^a_{n+1/2}|\phi\rangle =0$
for $n\geq 0$ and $a=1, 2$ (such a chiral primary state
$|\phi\rangle$ satisfies $L_0|\phi\rangle = T_0^3|\phi\rangle$).
From Eqs. (\ref{boson}), (\ref{fermion}) we then find the
supergravity elliptic genus to be
\begin{equation}
\chi^{\rm sugra}_{\rm NS}(\tau) = \cR(s= 2(1-i\varrho(\tau))) \cdot [\cR(s=
1-i\varrho(\tau))]^{{\mathcal C}(n_V, n_H, n_S)} \cdot \prod_{n\in {\mathbb Z}_+}
[\cR(s= (n+1)(1-i\varrho(\tau)))]^{-{\mathcal {\mathcal E}}},
\end{equation}
where the number ${\mathcal C}(n_V, n_H, n_S)$ depends on the
multiplets, and $\mathcal {\mathcal E}$ denotes the Euler number associated
with $CY_3$ manifold.

\section{Conclusions}

In this paper we have discussed how the modular and spectral functions of the
$AdS_3$-asymptotic geometry are intertwined with the quantum partition functions of 
gravity and of conformal field theory. The quantum corrections can be systematically included, by making use of the comparison with the result of conformal field theory via the existing $AdS_3/CFT_2$ correspondence. We have calculated the sub-leading corrections to the entropy of black hole (including a precise evaluation of the degeneracy prefactor).

The common link of all these examples is to be found, in our opinion, in an important feature of the theory of infinite dimensional Lie algebras, namely the modular properties of the characters (generating functions) of certain representations. The highest-weight modules of the affine Lie algebras underlie conformal field theories. 
The character of the highest-weight Virasoro module can be interpreted as the holomorphic part of the partition functions on the torus, for the corresponding conformal field theories. 
The quantum corrections for the three-dimensional gravity and the elliptic genus from supergravity states can be written in terms of spectral functions of hyperbolic geometry, providing holomorphically factorized results, spectral flow and a kind of modular invariance.
In many physical applications, quantum generating functions can be reproduced in terms of Selberg-type spectral functions. Therefore, the role of the unimodular group $SL(2; {\mathbb C})$ (and of the modular group $SL(2; {\mathbb Z})$) constitute a very clear manifestation of the link of all the above with hyperbolic three-geometry and its spectral functions.

\subsection*{Acknowledgements}

We wish to thank the referee for  useful comments and recommendations.   
We are  grateful  to Shahin Sheikh-Jabbari and Richard Szabo for discussions and suggestions.  Our special thanks go to Masud Chaichian for several helpful discussions,  for crucial remarks and suggestions. 
A. A. B. would like to acknowledge the Conselho Nacional
de Desenvolvimento Cient\'ifico e Tecnol\'ogico (CNPq, Brazil) and Funda\c cao Araucaria 
(Parana, Brazil) for financial support. The support of the Academy of Finland under the 
Projects No. 136539 and 140886 is gratefully acknowledged.

\end{document}